\documentclass[runningheads]{llncs}
\usepackage{graphicx}

\usepackage{amsmath}
\usepackage{amssymb}
\usepackage{bm}
\usepackage{subcaption}
\usepackage{color}
\usepackage{multirow}
\usepackage{booktabs}
\usepackage{hyperref}

\begin{document}
\title{Encoding Visual Attributes in Capsules for Explainable Medical Diagnoses}
\titlerunning{X-Caps: Explainable Medical Diagnoses}
%
\author{Rodney LaLonde\inst{1} \and
Drew Torigian\inst{2} \and
Ulas Bagci\inst{1}}
\authorrunning{R. LaLonde et al.}
%
\institute{Center for Research in Computer Vision, University of Central Florida \and 
Penn Medicine, University of Pennsylvania
}

\maketitle              
\begin{abstract}
Convolutional neural network based systems have largely failed to be adopted in many high-risk application areas, including healthcare, military, security, transportation, finance, and legal, due to their highly uninterpretable ``black-box'' nature. Towards solving this deficiency, we teach a novel multi-task capsule network to improve the explainability of predictions by embodying the same high-level language used by human-experts. Our explainable capsule network, \textbf{\textit{X-Caps}}, encodes high-level visual object attributes within the vectors of its capsules, then forms predictions based solely on these human-interpretable features. To encode attributes, \textit{X-Caps} utilizes a new routing sigmoid function to independently route information from child capsules to parents. Further, to provide radiologists with an estimate of model confidence, we train our network on a distribution of expert labels, modeling inter-observer agreement and punishing over/under confidence during training, supervised by human-experts' agreement. \textit{X-Caps} simultaneously learns attribute and malignancy scores from a multi-center dataset of over $1000$ CT scans of lung cancer screening patients. We demonstrate a simple 2D capsule network can outperform a state-of-the-art deep dense dual-path 3D CNN at capturing visually-interpretable high-level attributes and malignancy prediction, while providing malignancy prediction scores approaching that of non-explainable 3D CNNs. To the best of our knowledge, this is the first study to investigate capsule networks for making predictions based on radiologist-level interpretable attributes and its applications to medical image diagnosis. Code is publicly available at \href{https://github.com/lalonderodney/X-Caps}{https://github.com/lalonderodney/X-Caps}.

\end{abstract}


\section{Introduction} \label{sec:intro}

\begin{figure}
    \centering
    \begin{subfigure}{.4\textwidth}
        \centering
        \includegraphics[width=0.97\linewidth]{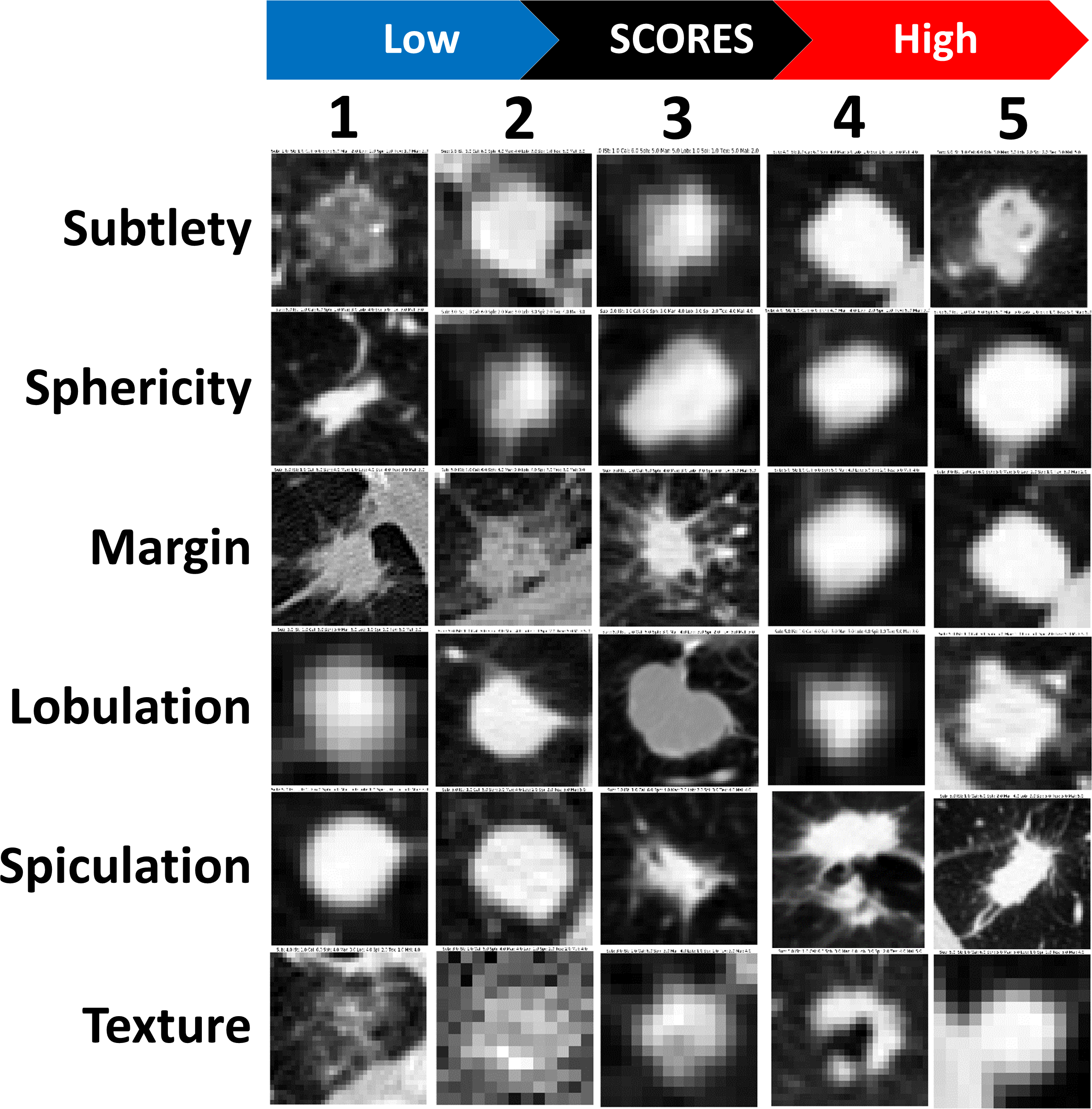}
        \caption{Lung nodules with high-level visual attribute scores as determined by expert radiologists. Scores were given from $1$ -- $5$ for six different visual attributes related to diagnosing lung cancer.}
        \label{fig:chars}
    \end{subfigure}
    \hspace{1em}
    \begin{subfigure}{.55\textwidth}
        \centering
        \includegraphics[width=.7\linewidth]{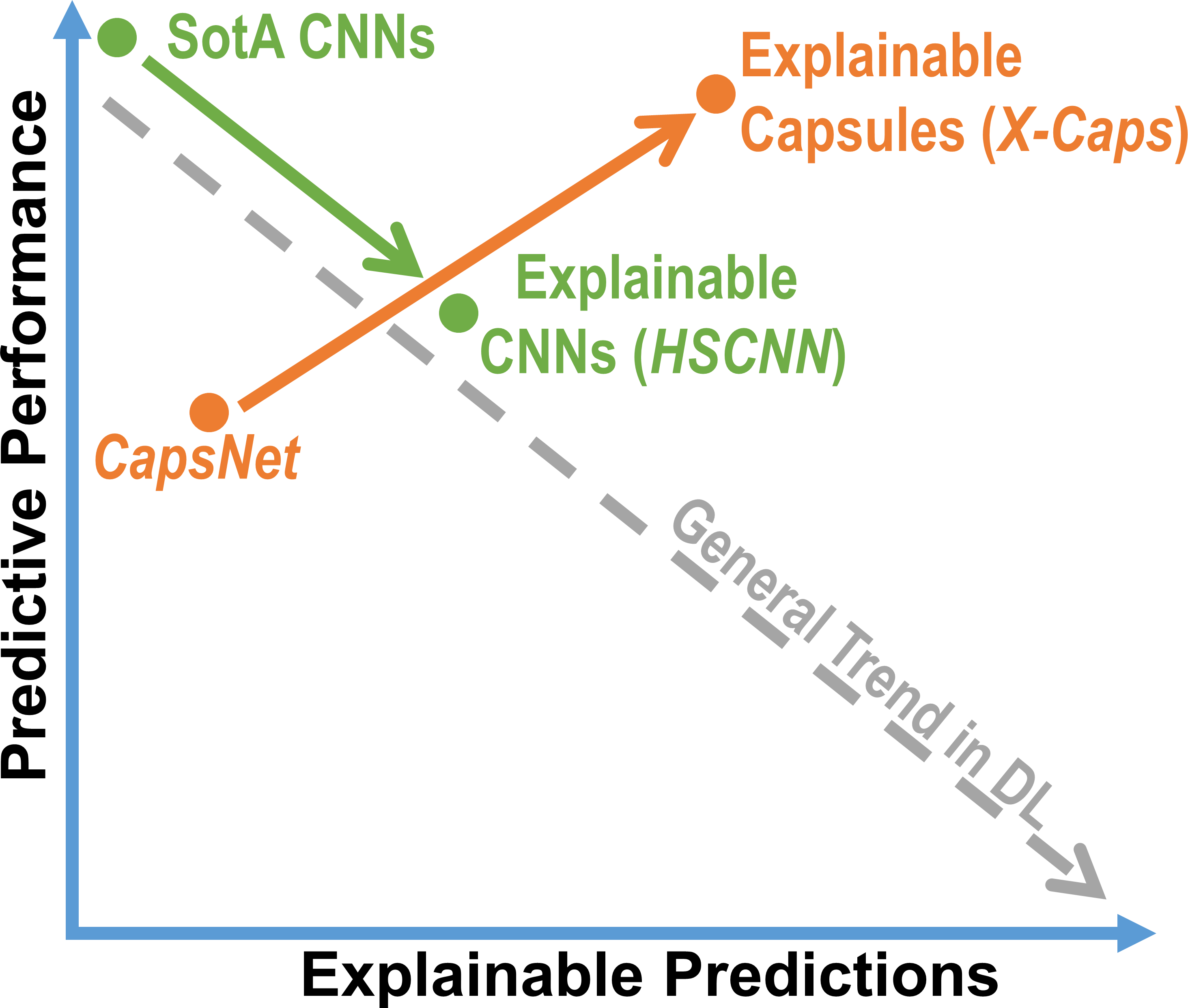}
        \caption{A symbolic plot showing the general trade-off between explainability and predictive performance in deep learning (DL)~\cite{bologna2003model,gilpin2018explaining,kuhn2013applied,rudin2019stop}. Our proposed \textit{X-Caps} rebuts the trend of decreasing performance from state-of-the-art (SotA) as explainability increases and shows it is possible to create more explainable models \textit{and} increase predictive performance with capsule networks.}
        \label{fig:trenddl}
    \end{subfigure}
    \caption{Encoding visual attributes (a) for explainable predictions (b).}
    \label{fig:fig1}
\end{figure}

In machine learning, predictive performance typically comes at the cost of \textit{interpretability}~\cite{bologna2003model,gilpin2018explaining,kuhn2013applied,rudin2019stop}. While deep learning (DL) has impact many fields, there exist several high-risk domains which have yet to be comparably affected: military, security, transportation, finance, legal, and healthcare among others, often citing a lack of interpretability as the main concern~\cite{bloomberg2018dont,lehnis2018can,polonski2018people}. As features become less \textit{interpretable}, and the functions learned more complex, model predictions become more difficult to explain (Fig.~\ref{fig:trenddl}). While some works have begun to press towards this goal of explainable DL, the problem remains largely unsolved.

\textbf{Interpretable vs. Explainable:} There has been a recent push in the community to move away from the \textit{post-hoc interpretations} of deep models and instead create explainable models from the outset \cite{rudin2019stop,shen2019interpretable}. Since the terms \textit{interpretable} and \textit{explainable} are often used interchangeably, we want to be explicit about our definitions for the purposes of this study. An \textit{explainable} model is one which provides explanations for its predictions \textit{at the human level} for a \textit{specific task}. An \textit{interpretable} model is one for which some conclusions can be drawn about the internals/predictions of the model; however, they are not explicitly provided by the model and are typically at a lower level. For example, in image classification, when a deep model predicts an image to be of a cat, saliency/gradient or other methods can attempt to \textit{interpret} the model/prediction. However, the model is not \textit{explaining} why the object in the image is a cat in the same way as a human. Humans classify objects based on a taxonomy of characteristics/attributes (\textit{e.g.} cat equals four legs, paws, whiskers, fur, etc.). If our goal is to create \textit{explainable} models, we should design models which explain their decisions using a similar set of ``attributes'' to humans.

\textbf{Why capsule networks?} Capsule networks differ from convolutional neural networks (CNNs) by replacing the scalar feature maps with vectorized representations, responsible for encoding information (\textit{e.g.} pose, scale, color) about each feature. These vectors are then used in a dynamic routing algorithm which seeks to maximize the agreement between lower-level predictions for the instantiation parameters (i.e. capsule vectors) of higher-level features. In their introductory work, a capsule network (\textit{CapsNet}) was shown to produce promising results on the MNIST data set; but more importantly, was able to encode high-level visually-interpretable features of digits (\textit{e.g.}  stroke thickness, skew, localized-parts) within the dimensions of its capsule vectors~\cite{sabour2017dynamic}.

\textbf{Lung cancer diagnosis with a multi-task capsule network:} In diagnosing the malignancy of lung nodules, similar to describing why an image of a cat is catlike, radiologists explain their predictions through the language of high-level visual attributes (i.e., radiographical interpretations): subtlety (sub), sphericity (sph), margin (mar), lobulation (lob), spiculation (spi), and texture (tex), shown in Fig.~\ref{fig:chars}, which are known to be predictive (with inherent uncertainty) of malignancy~\cite{hancock2016lung}. To create a DL model with this same level of radiographical interpretation, we propose a novel multi-task capsule architecture, called \textbf{\textit{X-Caps}}, for learning visually-interpretable feature representations within capsule vectors, then predicting malignancy based solely on these interpretable features. By supervising different capsules to embed specific visually-interpretable features, multiple visual attributes are learned simultaneously, with their weights being updated by both the radiologists visual interpretation scores as well as their contribution to the final malignancy score, regularized by the segmentation reconstruction error. Since these attributes are not mutually-exclusive, we introduce a new routing sigmoid function to independently route child capsules to parents. Further, to provide radiologists with an estimate of model confidence, we train our network on a distribution of expert labels, modeling inter-observer agreement and punishing over/under confidence during training, supervised by human-experts' agreement. 

We show even a relatively simple 2D capsule network can better capture high-level visual attribute information than the state-of-the-art deep dual-path dense 3D convolutional neural network (CNN) while also improving diagnostic accuracy, approaching that of even some black-box methods (\textit{e.g.,}~\cite{shen2015multi,shen2017multi}). \textit{X-Caps} simultaneously learns attribute and malignancy scores from a multi-center dataset of over $1000$ CT scans of lung cancer screening patients. \textbf{Overall, the contributions of this study are summarized as: }
\begin{enumerate}
    \item The first study to directly encode high-level visual attributes within the vectors of a capsule network to perform explainable image-based diagnosis \textit{at the radiologist-level}.
    \item Create a novel modification to the dynamic routing algorithm to independently route information from child capsules to parents when parent capsules are not mutually-exclusive.
    \item Provide a meaningful confidence metric with our predictions at test by learning directly from expert label distributions to punish network over/under confidence. Visual attribute predictions are verified at test via the reconstruction branch of the network.
    \item Demonstrate a simple 2D capsule network (\textit{X-Caps}) trained from scratch outperforming a state-of-the-art deep pre-trained dense dual-path 3D CNN at capturing visually-interpretable high-level attributes and malignancy prediction, while providing malignancy prediction scores approaching that of non-explainable 3D CNNs.
\end{enumerate}

\section{Related work} \label{sec:related}
The majority of work in explainable deep learning has focused around \textit{post hoc} deconstruction of already trained models (\textit{i.e.} interpretation). These approaches typically rely on human-experts to examine their results and attempt to discover meaningful patterns. Zeiler and Fergus \cite{zeiler2014visualizing} attached a deconvolutional network to network layers to map activations back to pixel space for visualizing individual filters and activation maps, while also running an occlusion-based study of which parts of the input contribute most to the final predictions. \textit{Grad-CAM}~\cite{selvaraju2017grad} is a popular approach which highlights the relative positive activation map of convolutional layers with respect to network outputs. \textit{InfoGAN}~\cite{chen2016infogan} separates noise from the ``latent code'', maximizing the mutual information between the latent representations and the image inputs, encoding concepts such as rotation, width, and digit type for MNIST. In a similar way, capsule networks encode visually-interpretable concepts such as stroke thickness, skew, rotation, and others~\cite{sabour2017dynamic}. 

A number of recent studies have proposed using \textit{CapsNet} for a variety of medical imaging classification tasks \cite{afshar2018brain,iesmantas2018convolutional,jimenez2018capsule,kandel2019su1741,mobiny2019automated,pal2018capsdemm,shen2018dynamic}. However, these methods nearly all follow the exact \textit{CapsNet} architecture, or propose minor modifications which present nearly identical predictive performance \cite{lalonde2020diagnosing,mobiny2018fast}; hence, it is sufficient to compare only with \textit{CapsNet} in reference to these works. 

In the area of lung nodule malignancy, many DL-based approaches have been proposed~\cite{shen2015multi,shen2017multi}, with further methods being developed with complicated post-processing techniques~\cite{ipmi}, curriculum learning methods~\cite{nibali2017pulmonary}, or gradient-boosting machines~\cite{zhu2018deeplung}. However, adding such techniques is beyond the scope of this study and would lead to an unwieldy enumeration of ablation studies necessary to understand the contributions between our proposed capsule architecture and such techniques. For a fair comparison in this study, we compare our method directly against \textit{CapsNet} and explainable CNN approaches. \textit{HSCNN}~\cite{shen2019interpretable} creates one of the first explainable methods, by designing a dense 3D CNN which first predicts visual attribute scores then predicts malignancy from those features. This decreased the overall performance as compared to other 3D networks~\cite{shen2017multi} but provided some explanations for the final malignancy predictions.

\section{Capsules for encoding visual attributes} \label{sec:methods}

\begin{figure*}[!ht]
\begin{center}
   \includegraphics[width=0.85\linewidth]{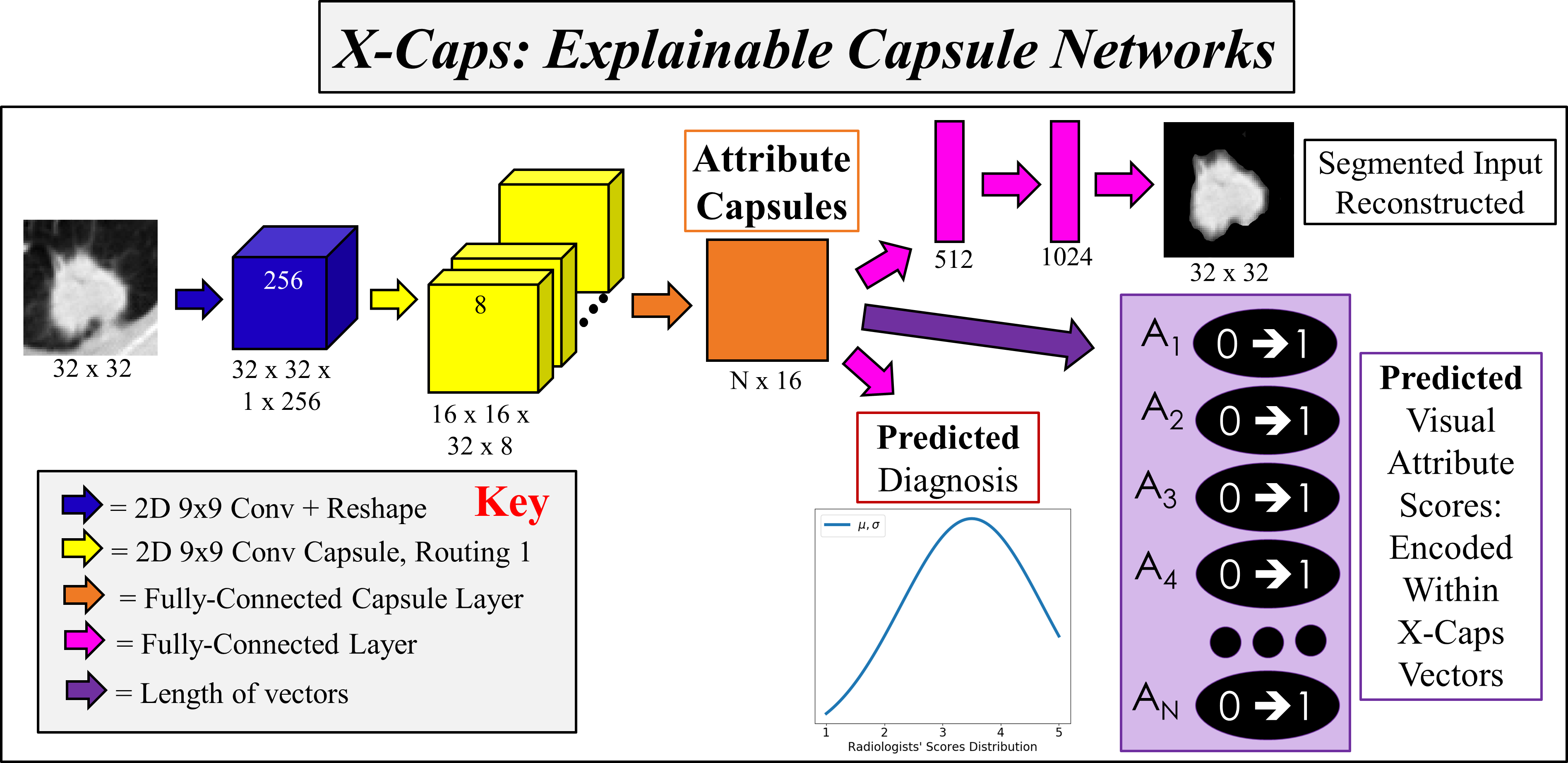}
\end{center}
   \caption{\textit{X-Caps}: Explainable Capsule Networks. The proposed network (1) predicts $N$ high-level visual attributes of the nodule, (2) segments the nodule and reconstruct the input image, and (3) diagnoses the nodule on a scale of 1 to 5 based on the visually-interpretable high-level features encoded in the X-Caps capsule vectors. The malignancy diagnosis branch is attempting to model the distribution of radiologists' scores in both mean and variance.}
\label{fig:xcaps}
\end{figure*}

Our approach, referred to as \textit{explainable capsules}, or \textit{X-Caps}, was designed to remain as similar as possible to our control network, \textit{CapsNet}, while allowing us to have more control over the visually-interpretable features learned. \textit{CapsNet} already showed great promise when trained on the MNIST data set for its ability to model high-level visually-interpretable features. With this study, we examine the ability of capsules to model \textit{specific} visual attributes within their vectors, rather that simply hoping these are learned successfully in the more challenging lung nodule data. As shown in Figure \ref{fig:xcaps}, \textit{X-Caps} shares a similar overall structure as \textit{CapsNet}, with the major differences being the addition of the supervised labels for each of the \textit{X-Caps} vectors, the fully-connected layer for malignancy prediction, the reconstruction regularization also performing segmentation, and the modifications to the dynamic routing algorithm.

The first layer of our proposed network is a 2D convolutional layer which extracts low-level features. Next, we form our primary capsules of $32$ capsule types with $8D$ vector capsules. Following this, we form our attribute capsules using a fully-connected capsule layer whose output is $N$ $16D$ capsule types, one for each of the visual-attributes we want to predict. Unlike \textit{CapsNet} where each of the parent capsules were dependant on one another (e.g. if the prediction is the digit $5$ it cannot also be a $3$), our parent capsules are not mutually-exclusive of each other (i.e. a nodule can score high or low in each of the attribute categories). For this reason, we needed to modify the dynamic routing algorithm presented in \textit{CapsNet} to accommodate this significant difference. The key change is the ``routing softmax'' employed by \textit{CapsNet} forces the contributions of each child to send their information to parents in a manner which sums to one, which in practice effectively makes them ``choose'' a parent to send their information to. However, when computing prediction vectors for independent parents, we want a child to be able to contribute to all parent capsules for attributes which are present in the given input. With that motivation, the specific algorithm, which we call ``routing sigmoid'', is computed as 
\begin{equation}\label{eq:sigmoid}
r_{i, j} = \frac{\exp(b_{i, j})}{\exp(b_{i, j}) + 1}, 
\end{equation}
where $r_{i, j}$ are the routing coefficients determined by the dynamic routing algorithm for child capsule $i$ to parent capsule $j$ and the initial logits, $b_{i, j}$ are the prior probabilities that the prediction vector for capsule $i$ should be routed to parent capsule $j$. Note the prior probabilities are initially set to $1$ rather than $0$ as in \textit{CapsNet}, otherwise no routing could take place. The rest of the dynamic routing procedure follows the same as in \cite{sabour2017dynamic}.

\textbf{Predicting malignancy from visually-interpretable capsule vectors:} 
In order to predict malignancy scores, we attach a fully-connected layer to our attribute capsules with output size equal to the range of scores. We wish to emphasize here, our final malignancy prediction is coming solely from the vectors whose magnitudes represent \textit{visually-interpretable} feature scores. Every malignancy prediction score has a set of weights connected to the high-level attribute capsule vectors, and the activation from each tells us the exact contribution of the given visual attribute to the final malignancy prediction for that nodule. Unlike previous studies which look at the importance of these attributes on a global level, our method looks at the importance of each visual attribute in relation to a specific nodule being diagnosed. To verify the correctness of our attribute modeling, we reconstruct the nodules while varying the dimensions of the capsule vectors to ensure the desired visual attributes are being modeled. At test, these reconstructions give confidence that the network is properly capturing the attributes, and thus the scores can be trusted. Confidence in the malignancy prediction score, in addition to coming solely from these trusted attributes, is provided via an uncertainty modeling approach.

Previous works in lung nodule classification follow the same strategy of averaging radiologists' scores for visual attributes and malignancy, and then either attempt to regress this average or performing binary classification of the average as below or above $3$. To better model the uncertainty inherently present in the labels due to inter-observer variation, we propose a different approach: we attempt to predict the \textit{distribution} of radiologists' scores. Specifically, for a given nodule where we have at minimum three radiologists' score values for each attribute and for malignancy prediction, we compute the mean and variance of those values and fit a Gaussian function to them, which is in turn used as the ground-truth for our classification vector. Nodules with strong inter-observer agreement produce a sharp peak, in which case wrong or unsure (\textit{i.e.,}  low confidence score) predictions are severely punished. Likewise, for low inter-observer agreement nodules, we expect our network to output a more spread distribution and it will be punished for strongly predicting a single class label. This proposed approach allows us to model the uncertainty present in radiologists' labels in a way that no previous study has and provide a meaningful confidence metric at test time to radiologists.

\textbf{Loss and regularization:} As in \textit{CapsNet}, we also perform reconstruction of the input as a form of regularization. However, we extend the idea of regularization to perform a pseudo-segmentation, similar in nature to the reconstruction used by \cite{lalonde2018capsules,lalonde2020capsules}. Whereas in true segmentation, the goal is to output a binary mask of pixels which belong to the nodule region, in our formulation we attempt to reconstruct only the pixels which belong to the nodule region, while the rest are mapped to zero. More specifically, we formulate this loss as
\begin{equation}
    \mathcal{L}_r = \frac{\gamma}{H \times W}\sum_x^W\sum_y^H\|R^{x,y} - O_r^{x,y}\|\textrm{, with }
    R^{x,y} = I^{x,y}\times S^{x,y} \mid S^{x,y} \in \{0,1\}\textrm{,}
    \label{eq:maskedloss}
\end{equation}
where $\mathcal{L}_r$ is the supervised loss for the reconstruction regularization, $\gamma$ is a weighting coefficient for the reconstruction loss, $R^{x,y}$ is the reconstruction target pixel, $S^{x,y}$ is the ground-truth segmentation mask value, and $O_r^{x,y}$ is the output of the reconstruction network, at pixel location $(x,y)$, respectively, and $H$ and $W$ are the height and width, respectively, of the input image. This adds another task to our multi-task approach and an additional supervisory signal which can help our network distinguish visual characteristics from background noise. The malignancy prediction score, as well as each of the visual attribute scores also provide a supervisory signal in the form of
\begin{equation}
    \mathcal{L}_a = \sum_n^N\alpha^n\|A^n - O_a^n\|\textrm{, } 
    \mathcal{L}_m = \beta \sum_{x \in X}\varepsilon(\bm{O}_m)\log\Bigg(\frac{\varepsilon(\bm{O}_m)}{\frac{1}{\sqrt{2\pi \sigma^2}}\exp\big(\frac{-(x-\mu)^2}{2\sigma^2}\big)}\Bigg) \textrm{,}
    \label{eq:attrloss}
\end{equation}
where $\mathcal{L}_a$ is the combined loss for the visual attributes, $A^n$ is the average of the attribute scores given by at minimum three radiologists for attribute $n$, $N$ is the total number of attributes, $\alpha^n$ is the weighting coefficient placed on the $n^{th}$ attribute, $O_a^n$ is the network prediction for the score of the $n^{th}$ attribute, $\mathcal{L}_m$ is a KL divergence loss for the malignancy score, $\beta$ is the weighting coefficient for the malignancy score, $\mu$ and $\sigma$ are the mean and variance of radiologists' scores, and $\varepsilon = \exp(O_m^i)/\sum_{j=1}^N\exp(O_m^j)$ is the softmax over the network malignancy prediction vector $\bm{O}_m = \{O_m^1, ..., O_m^N \}$. In this way, the overall loss for \textit{X-Caps} is simply $\mathcal{L} = \mathcal{L}_m + \mathcal{L}_a + \mathcal{L}_r$. For simplicity, the values of each $\alpha^n$ and $\beta$ are set to $1$, and $\gamma$ is set to $0.005 \times 32 \times 32 = 0.512$.

\section{Experiments, results, limitations, and ablations} \label{sec:exp}

Experiments we performed on the LIDC-IDRI data set~\cite{LIDC}. Five-fold stratified cross-validation was performed, with $10\%$ of each training set used for validation and early stopping. \textit{X-Caps} was trained with a batch size of $16$ using Adam with an initial learning rate of $0.02$ reduced by a factor of $0.1$ after validation loss plateau. Consistent with the literature, nodules of mean radiologists' score $3$ were removed (leaving $646$ benign and $503$ malignant nodules) and predictions were considered correct if within $\pm1$ of the radiologists' classification \cite{ipmi,tumornet}. The results summarized in Table~\ref{table:diagnosis} illustrate the prediction of visual attributes with the proposed \textit{X-Caps} in comparison with an adapted version of \textit{CapsNet}, a deep dense dual-path 3D explainable CNN (\textit{HSCNN}~\cite{shen2019interpretable}), and two state-of-the-art non-explainable methods which do not have extra post-processing or learning strategies. Compared methods results are from the original reported works.

\begin{table*}[htbp!]
\begin{center}
\caption{Prediction accuracy of visual attributes with capsule networks. Dashes (-) represent values which the given method could not produce. \textit{X-Caps} outperforms the state-of-the-art explainable method (\textit{HSCNN}) at attribute modeling (the main goal of both studies), while also producing higher malignancy prediction scores, approaching state-of-the-art non-explainable methods performance.
\label{table:diagnosis}}
\begin{tabular}{|l||c|c|c|c|c|c||c|}
\hline
 & \multicolumn{6}{|c|}{\textbf{Attribute Prediction Accuracy \%}} & \textbf{Malig-}\\
& sub & sph & mar & lob & spi & tex & \textbf{nancy} \\
\hline
\underline{\textbf{Non-Explainable Methods}} & & & & & & & \\ 
3D Multi-Scale + RF~\cite{shen2015multi} & - & - & - & - & - & - & $86.84$ \\
3D Multi-Crop~\cite{shen2017multi} & - & - & - & - & - & - & $87.14$ \\
\textit{CapsNet}~\cite{sabour2017dynamic} & - & - & - & - & - & - & $77.04$ \\
\hline
\underline{\textbf{Explainable Methods}} & & & & & & & \\ 
3D Dual-Path \textit{HSCNN}~\cite{shen2019interpretable} & $71.9$ & $55.2$ & $72.5$ & - & - & $83.4$ & $84.20$ \\
\textbf{Proposed \textit{X-Caps}} & $\bm{90.39}$ & $\bm{85.44}$ & $\bm{84.14}$ & $\bm{70.69}$ & $\bm{75.23}$ & $\bm{93.10}$ & $\bm{86.39}$ \\
\hline
\end{tabular}
\end{center}
\vspace{-0.5cm}
\end{table*}

Our results show that a \textit{X-Caps} has the ability to model visual attributes far better than \textit{HSCNN} while also achieving better malignancy prediction. Further, we wish to emphasize the significance of \textit{X-Caps} providing increased predictive performance \textit{and} explainability over \textit{CapsNet}. This goes against the assumed trend in DL, illustrated with a symbolic plot in Figure~\ref{fig:trenddl}, that explainability comes at the cost of predictive performance, a trend we observe with \textit{HSCNN} being outperformed by less powerful (\textit{i.e.} not dense or dual-path) but non-explainable 3D CNNs~\cite{shen2015multi,shen2017multi}. While \textit{X-Caps} slightly under-performs the best non-explainable models, it is reasonable to suspect that future research into more powerful 3D capsule networks would allow explainable capsules to surpass these methods; we hope this study will promote such future works. 

As two limitations of our work, we did not tune the weight balancing terms between the different tasks and further investigation could lead to superior performance. Also, we found capsule networks can be somewhat fragile; often random initializations failed to converge to good performance. However, this might be due to the small/shallow network size and its relation to the Lottery Ticket Hypothesis \cite{frankle2018lottery} rather than anything specific to capsules.


\textbf{Ablation studies:} To analyze the impact of each component of our proposed approach, we performed ablation studies for: (1) learning the distribution of radiologists' scores rather than attempting to regress the mean value of these scores, (2) removing the reconstruction regularization from the network, and (3) performing our proposed ``routing sigmoid'' over the original ``routing softmax'' proposed in \cite{sabour2017dynamic}. The malignancy prediction accuracy for each of these ablations is (1) $83.09\%$, (2) $80.30\%$, and (3) $80.69\%$, respectively, as compared to the proposed model's accuracy of $86.39\%$. This shows retaining the agreement/disagreement information among radiologists proved useful, the reconstruction played a role in improving the network performance, and our proposed modifications to the dynamic routing algorithm were necessary for passing information from children to parents when the parent capsule types are independent. 

\section{Discussions and concluding remarks} \label{sec:disc}

Available studies for explaining DL models, typically focus on \textit{post hoc} interpretations of trained networks, rather than attempting to build-in explainability. This is the first study for directly learning an interpretable feature space by encoding high-level visual attributes within the vectors of a capsule network to perform explainable image-based diagnosis. We approximate visually-interpretable attributes through individual capsule types, then predict malignancy scores directly based only on these high-level attribute capsule vectors, in order to provide malignancy predictions with explanations \textit{at the human-level}, in the same language used by radiologists. Our proposed multi-task explainable capsule network, \textit{X-Caps}, successfully approximated visual attribute scores better than the previous state-of-the-art explainable diagnosis system, while also achieving higher diagnostic accuracy. We hope our work can provide radiologists with malignancy predictions which are explained via the same high-level visual attributes they currently use, while also providing a meaningful confidence metric to advise when the results can be more trusted, thus allowing radiologists to quickly interpret and verify our predictions. Lastly, we believe our approach should be applicable to any image-based classification task where high-level attribute information is available to provide explanations about the final prediction.

%
%

\bibliographystyle{splncs04}
\bibliography{main}

\end{document}